\documentclass[preprint,aps,showpacs,nofootinbib]{revtex4-1}

\usepackage{amssymb}
\usepackage{bm}

\def\bea{\begin{eqnarray}}
\def\eea{\end{eqnarray}}
\def\bec{\begin{center}}
\def\ec{\end{center}}

\def\beq{\begin{equation}}
\def\eeq{\end{equation}}

\newcommand{\GeV}{\mathinner{\mathrm{GeV}}}
\newcommand{\TeV}{\mathinner{\mathrm{TeV}}}

\newcommand{\axino}{{\tilde{a}}}

\newcommand{\planck}{\mathrm{Pl}}

\begin{document}

\begin{center}
{\Large \bf  The $\mu$-problem and axion in gauge mediation}
\end{center}

 \vspace*{5mm} \noindent
\centerline{\bf Kiwoon Choi${}^{a}$,  Eung Jin Chun${}^{b}$,  Hyung
Do Kim${}^{c}$, Wan Il Park${}^{b}$, Chang Sub Shin${}^{a}$} \vskip
1cm \centerline{\em ${{}^{a}}$ Department of Physics, Korea Advanced
Institute of Science and Technology} \centerline{\em Daejeon
305-701, Korea}
\vskip 5mm \centerline{\em ${{}^{b}}$ Korea Institute for
Advanced Study, Hoegiro 87, Dongdaemun-gu} \centerline{\em
Seoul 130-722, Korea}
\vskip 5mm \centerline{\em ${{}^{c}}$ School of
Physics and Astronomy, Seoul National University} \centerline{\em
Seoul 151-747, Korea}

\vskip .9cm

\centerline{\bf Abstract} \vskip .3cm

We revisit the idea of generating the Higgs  $\mu$  parameter
through a spontaneously broken Peccei-Quinn  (PQ) symmetry in
gauge-mediated supersymmetry breaking scenario.
For the messenger scale of gauge mediation  higher than the PQ
scale,
the setup naturally generates  $\mu\sim m_{\rm soft}$ and the Higgs
soft parameter $B\lesssim {\cal O}(m_{\rm soft})$ with the CP phase
of $B$ aligned to the phase of gaugino masses, while giving the PQ
scale $v_{PQ}\sim \sqrt{m_{\rm soft}\Lambda}$, where $m_{\rm soft}$
denotes the gauge-mediated gaugino or sfermion masses and $\Lambda$
is the cutoff scale which can be identified as the Planck scale or
the GUT scale.
The PQ sector of the model results in distinctive cosmology
including a late thermal inflation. We discuss the issue of dark
matter and baryogenesis in the resulting thermal inflation scenario,
and find that a right amount of gravitino dark matter can be
produced  together with a successful Affleck-Dine leptogenesis, when
the gravitino mass $m_{3/2}={\cal O}(100)$ keV.


\vskip .3cm


\newpage

\section{introduction}

Weak scale supersymmetry (SUSY) is one of the most attractive
candidates for new physics beyond the standard model (SM) at the TeV
scale  \cite{susy}.
It provides an appealing  solution to
the gauge hierarchy problem, and also the successful unification of
gauge couplings at the scale $M_{GUT}\sim 2\times 10^{16}$ GeV. On
the other hand, the absence of unacceptably large flavor or CP
violations requires a rather special type of supersymmetry breaking
which yields flavor and CP conserving soft terms.
 Supersymmetry breaking
through gauge mediation \cite{Giudice:1998bp}
provides flavor conserving (possibly CP conserving also) soft terms
in a natural manner  as the structure of soft terms is determined
mostly by the SM gauge interactions. One potential difficulty of
gauge mediation mechanism is the generation of the Higgs $\mu$ and $B$
parameters having a right size for the electroweak symmetry
breaking. If the Higgs sector communicates with the SUSY breaking
sector  to generate $\mu\sim m_{\rm soft}$, where $m_{\rm soft}$
denotes the gaugino and sfermion masses in gauge mediation, one
often finds $B\sim 8\pi^2 m_{\rm soft}$, which is too large to
achieve a successful electroweak symmetry breaking. There have been
many attempts to solve the $\mu$ problem in gauge mediation,
including those in Ref.~\cite{mu-gmsb}.


As was noticed in the original work of Kim and Nilles
\cite{Kim:1983dt},  a satisfactory solution of the $\mu$ problem
should provide a theoretical reasoning for the absence of the bare
$\mu$ term with $\mu\sim \Lambda$, as well as a dynamical mechanism
to generate $\mu\sim m_{\rm soft}$ together with $B\sim m_{\rm
soft}$ at the weak scale, where $\Lambda$ denotes the cutoff scale
of the model
 which can be taken as either the reduced Planck scale
$M_{Pl}\sim 2\times 10^{18}$ GeV or the GUT scale $M_{GUT}\sim
2\times 10^{16}$ GeV.  It was further noticed in Ref.~\cite{Kim:1983dt}
that the Peccei-Quinn (PQ) symmetry solving the strong CP problem
might play a crucial role for the $\mu$ problem as well. The
$U(1)_{PQ}$ symmetry might forbid the bare $\mu$ term, while
allowing the following non-renormalizable  term in the
superpotential \bea \label{mu-term} \frac{1}{\Lambda}X^2H_uH_d, \eea
where $X$ is a PQ charged SM singlet field  whose vacuum value
breaks $U(1)_{PQ}$ spontaneously.
 If the PQ sector
of the model couples to the SUSY breaking sector  to stabilize $X$
at \bea v_{PQ}\,\equiv\,\langle X\rangle \,\sim\, \sqrt{m_{\rm
soft}\Lambda},\eea and the $F$-component of the stabilized $X$
satisfies ${F^X}/{X}\lesssim {\cal O}(m_{\rm soft})$, the resulting
$\mu$ and $B$ (at the weak scale) have a right size for successful
electroweak symmetry breaking.

The Kim-Nilles mechanism was discussed originally in the context of
gravity mediation with $m_{\rm soft}\sim m_{3/2}$
\cite{Kim:1983dt,CKN,Murayama:1992dj}.
Later it was realized  that  the mechanism
can be implemented also in gauge mediation  \cite{Chun:1997mh}.
However, the specific models discussed in \cite{Chun:1997mh} involve
$U(1)_{PQ}$ which is assumed to be an $R$-symmetry. In such models,
the nonzero vacuum value of the superpotential, which is required to
tune the cosmological constant vanish, should appear as a
consequence of the spontaneous breakdown of $U(1)_{PQ}$, and this
makes a complete realization of the setup  quite complicate.
In this paper, we revisit the Kim-Nilles mechanism to generate $\mu$
and $B$ in gauge mediation, while focusing on the case that
$U(1)_{PQ}$ is not an $R$-symmetry, but an ordinary anomalous global
symmetry.
It is noticed that such class of models can have a distinctive
cosmological feature such as a late thermal inflation triggered by
the PQ sector \cite{Lyth:1995hj}.  We then need a late baryogenesis
after thermal inflation as well as a mechanism to produce a right
amount of dark matter. We find that a right amount of gravitino dark
matter can be produced after thermal inflation when $v_{PQ}={\cal
O}(10^9 - 10^{10})$ GeV and $m_{3/2}={\cal O}(100)$ keV.
With the
nonrenormalizable term (\ref{mu-term}) and also the seesaw term  for
light Majorana neutrino masses,  the model can accommodate also a
successful Affleck-Dine leptogenesis proposed in
\cite{Stewart:1996ai,Jeong:2004hy}.

This paper is organized as follow. In section 2, we discuss the
Kim-Nilles mechanism to generate $\mu\sim m_{\rm soft}$ and
$B\lesssim {\cal O}(m_{\rm soft})$  with a spontaneously broken
$U(1)_{PQ}$ in gauge mediation. Section 3 discusses the cosmological
aspects of the model, including the mechanisms to generate the right
amount of dark matter and baryon asymmetry, and the conclusion will
be given in section 4.

\section{The Kim-Nilles mechanism in gauge mediation}

In (minimal) gauge mediation scenario,  SUSY breaking is mediated by
 SM gauge-charged messengers $\Phi,\Phi^c$ which couple to  SUSY
breaking field $Z=M+F\theta^2$ in the superpotential. Then sfermions
and gauginos in the minimal supersymmetric standard model (MSSM) get
soft SUSY breaking masses  \bea m_{\rm soft}\sim
\frac{g^2}{16\pi^2}\frac{F}{M}\eea through the loops involving the
messenger fields $\Phi,\Phi^c$.
In order to implement the Kim-Nilles mechanism to generate the $\mu$
term, we introduce additional SM singlet but PQ charged superfields
which break $U(1)_{PQ}$ spontaneously, and also extra vector-like
quark superfields\footnote{To keep the successful unification of
gauge couplings in the MSSM, these extra vector-like quarks can be
extended to form a full GUT multiplet.} which have a Yukawa coupling
to the $U(1)_{PQ}$-breaking fields.
 With
$U(1)_{PQ}$, one can forbid renormalizable superpotential term of
the $U(1)_{PQ}$-breaking fields, while allowing a nonrenormalizable
term suppressed by the cutoff scale $\Lambda$ of the model. Then due
to the SUSY breaking effects mediated through the Yukawa coupling to
extra quark superfields, the $U(1)_{PQ}$-breaking fields are
destabilized at the origin. On the other hand, the supersymmetric
scalar potential originating from the nonrenormalizable
superpotential prevents the runaway of the $U(1)_{PQ}$-breaking
fields, and stabilize them at an intermediate scale $v_{PQ}\sim
\sqrt{m_{\rm soft}\Lambda}$. With $\Lambda$ presumed to be the GUT
scale or the Planck scale, this scenario naturally generates a QCD
axion scale $v_{PQ}={\cal O}(10^9-10^{11})$ GeV, as well as a
correct size of $\mu\sim v_{PQ}^2/\Lambda\sim m_{\rm soft}$ via the
Kim-Nilles mechanism. Furthermore, in this setup one can easily
obtain the Higgs $B$ parameter at the weak scale which is (at most)
comparable to $m_{\rm soft}$ and has a CP phase aligned to the phase
of gaugino masses.



As a specific model to realize this scenario, we consider the
superpotential \bea \label{super}
W & = & y_u Q H_u u^c + y_d Q H_d d^c + y_e L H_d e^c + \frac{y_\nu}{M_N}LH_uLH_u,\nonumber \\
&& + \lambda_X X \Psi \Psi^c + \frac{\kappa_1}{6\Lambda} X^3 Y + \frac{\kappa_2}{2\Lambda} X^2 H_u H_d \nonumber \\
&&+ \lambda_Z Z \Phi \Phi^c \eea where the first line denotes the
usual Yukawa couplings between the Higgs  fields and the quarks
and/or leptons, including the term for small neutrino masses which
might be generated by the seesaw mechanism \cite{Mohapatra:2005wg}
with a right handed neutrino mass $M_N$ far above the weak scale.
Here the flavor indices are omitted, and $y_u,y_d,y_e$ and $y_\nu$
should be understood as $3 \times 3$ matrices. As we will see, the
above model has a variety of interesting cosmological features,
including a late thermal inflation associated with the PQ phase
transition in the early universe. Although we do not specify the
origin of $M_N$ here, an interesting possibility is  that $M_N$ is
generated as a consequence of $U(1)_{PQ}$ breaking, which would give
$M_N \sim \langle X\rangle$, so that the seesaw scale is identified
as the PQ scale\footnote{Of course, then the Yukawa couplings
between $H_u$ and the left and right handed neutrinos should have
appropriately small values  to produce the observed neutrino
mass-square differences and mixing angles.} \cite{kang}. As we will
see in the next section, such setup can be useful also for a
successful Affleck-Dine leptogenesis after thermal inflation.

 The second line of the superpotential (\ref{super})
is the PQ sector generating the Higgs $\mu$ parameter through the
Kim-Nilles mechanism, while providing a QCD axion to solve the
strong CP problem \cite{strongcp}. The third line is for the minimal
gauge mediation of SUSY breaking, where $Z$  is the SUSY breaking
field with \bea \langle Z\rangle=M+F\theta^2.\eea
Note that we can always make $\lambda_X, \kappa_1,\kappa_2$ and
$\lambda_Z\langle Z\rangle$ all real and positive through
appropriate field redefinitions, and we will take such field basis
in the following discussion.
 To
be specific, we also assume  that the cutoff scale $\Lambda$ is
around the GUT scale \bea \Lambda\sim M_{GUT}=2\times 10^{16}\,\,
{\rm GeV}.\eea
Although different choice of $\Lambda$ would change the value of the
PQ scale, the $\mu$ parameter obtained by the Kim-Nilles mechanism
is independent of $\Lambda$ and always of the order of $m_{\rm
soft}$ as long as the dimensionless parameters $\kappa_1$ and
$\kappa_2$ have a similar size. Note that the superpotential
(\ref{super}) takes the most general form (up to ${\rm dim}=4$
terms) allowed by the SM gauge symmetries, $R$-parity and
$U(1)_{PQ}$, where the $U(1)_{PQ}$ charges are given as follows:
\[
\begin{array}{|c|c|c|c|c|c|c|c|c|c|c|}
\hline
{\rm Field}  & Z & X & Y & H_u & H_d & Q u^c & Q d^c & L e^c & \Psi \Psi^c & \Phi \Phi^c\\
  \hline
{\rm PQ \ charge} & 0 & 1 & -3 & -1 & -1 & 1 & 1 & 1 & -1 & 0 \\
\hline
\end{array}
\]


Let us now discuss the vacuum configuration of the
$U(1)_{PQ}$-breaking fields $X$ and $Y$.
As the gravitino mass is much smaller than the weak scale, we can
safely ignore the supergravity effects. Then the scalar potential of
$X,Y$ can be well approximated by the global SUSY potential
including the soft SUSY breaking terms induced by radiative
corrections: \bea \label{potential}V(X,Y) & = & V_{\rm soft} (X) +
\frac{\kappa_1^2}{36\Lambda^2} |X|^6 + \frac{\kappa_1^2}{4\Lambda^2}
|X|^4 |Y|^2, \eea where \bea V_{\rm soft} (X,Y) & = & m_X^2 |X|^2 +
m_Y^2 |Y|^2 + \left(A_{\kappa_1} \frac{\kappa_1}{6\Lambda} X^3 Y+
{\rm h.c.}\right). \nonumber \eea If the messenger scale $M_\Phi$ of
gauge mediation is above the PQ threshold scale \bea M_\Phi \equiv
\lambda_Z \langle Z\rangle \gtrsim \lambda_X \langle X\rangle,\eea
which is in fact necessary to generate $\mu\sim m_{\rm soft}$
independently of the value of $\Lambda$, the soft mass $m_X$ at
scales below $M_\Phi$ is generated mostly by the renormalization
group (RG) running triggered by the Yukawa coupling
$\lambda_XX\Psi\Psi^c$.
 The RG equation for $m_X^2$ is   given by
\bea \frac{d m_X^2}{d \ln \mu^2} & = & \frac{3 \lambda_X^2}{16\pi^2}
\left( m_{\tilde{\Psi}}^2 + m_{\tilde{\Psi}^c}^2 + m_X^2 + |A_{X
\Psi \Psi^c}|^2\right), \eea where the factor 3 is the color factor,
and  $m_{\tilde{\Psi}},m_{\tilde{\Psi}^c}$ and $A_{X \Psi \Psi^c}$
are   the gauge-mediated soft scalar masses and trilinear scalar
coupling, respectively,  for the squark components of $\Psi,\Psi^c$.
 As $m_X$ and $A_{X\Psi\Psi^c}$ at the messenger scale are negligible compared to $m_{\tilde\Psi}\sim m_{\rm soft}$,
the soft mass $m_X$  at lower renormalization point $\langle X
\rangle$ is determined as \bea \label{mx} m_X^2 (|X|) & \simeq & -
(m_{\tilde{\Psi}}^2 + m_{\tilde{\Psi}^c}^2) \frac{3
\lambda_X^2}{8\pi^2} \ln \frac{M_\Phi}{|X|}, \eea where
we have ignored higher powers of $\frac{1}{8\pi^2}\ln ({M_\Phi}/{|X
|})$ with the assumption that $M_\Phi$ is not so far above $\langle
X\rangle$. Since  the PQ breaking scale $\langle X\rangle$ is
constrained to be of  ${\cal O}(10^{9}-10^{12})$ GeV, while the
messenger scale should be lower than ${\cal O}(10^{15})$ GeV in
order for the gauge mediation to give dominant contribution to soft
terms, the value of $\ln ({M_\Phi}/{| X |})$ can not be so large,
and therefore our assumption is justified.

In our approximation, $m_{\tilde{\Psi}}$ and $m_{\tilde{\Psi}^c}$
in (\ref{mx}) can be regarded as the soft squark masses  at
 the messenger scale $M_\Phi$, which are given by
   \bea m_{\tilde\Psi}^2=m_{\tilde{\Psi}^c}^2 = \frac{8N_\Phi}{3}
\left| \frac{g_s^2}{16\pi^2}\frac{F}{M} \right|^2\,\simeq\, m_{\rm
soft}^2, \eea where
$N_\Phi$ is the number of messenger pairs in the fundamental
representation.  Minimizing the scalar potential (\ref{potential})
with the tachyonic $m_X^2$ given by (\ref{mx}), we find \bea
\label{vpq} v_{PQ}^2\,\equiv \,\langle |X|^2\rangle \,\simeq\,
\frac{3\lambda_X\sqrt{\ln
(M_\Phi/v_{PQ})}}{\pi\kappa_1}m_{\tilde{\Psi}}\Lambda\,=\,{\cal
O}\left(m_{\rm soft}\Lambda\right)\eea and
 the resulting Higgs $\mu$ parameter  \bea \mu \,\simeq\,
\frac{3\kappa_2\lambda_X\sqrt{\ln
(M_\Phi/v_{PQ})}}{2\kappa_1\pi}m_{\tilde{\Psi}}\,=\,{\cal
O}\left(m_{\rm soft}\right),\eea where  we assumed that
$\lambda_X,\kappa_1$ and $\kappa_2$ are all of order unity
 for the order of magnitude
estimate in the last step.
Note that $\mu$ is independent of the precise value of the cutoff
scale $\Lambda$, while the PQ scale  has a mild dependence on
$\Lambda$.

The VEV of $X$ in (\ref{vpq}) generates an effective mass of $Y$
through the term $\propto |X|^4|Y|^2$ in the scalar potential:\bea
\frac{\kappa_1^2}{4\Lambda^2}\langle |X|^4\rangle
|Y|^2=3|m_X|^2|Y|^2.\eea It also generates an effective tadpole of
$Y$ through the $A$ term $\propto X^3Y$ in the scalar potential,
which is generated by the RG evolution of the wavefunction factor of
$X$. We then find \bea \frac{A_{\kappa_1} (|X|)}{m_{\tilde{\Psi}}} &
\simeq & - \frac{3\sqrt{3N_{\Phi}}\lambda_X^2}{4\sqrt{2}\pi^2}
\frac{g_s^2}{8\pi^2}  \left(\ln \frac{M_\Phi}{|X|}\right)^2,
\label{eq:A} \eea and therefore \bea \frac{\langle |Y|
\rangle}{\langle |X| \rangle} & =
&\frac{|A_{\kappa_1}|}{3\sqrt{3}|m_X|} \simeq \frac{\sqrt{
N_\Phi}}{2\sqrt{6}}\frac{g_s^2\lambda_X}{8\pi^3}\left(\ln \frac{M_\Phi}{\langle |X| \rangle}\right)^{3/2}.
\eea With the above results, one can  compute the Higgs
$B$-parameter around the messenger scale, which is given by \bea
B(M_\Phi)\equiv
\left.\frac{B\mu}{\mu}\right|_{M_\Phi}=2\left(\frac{F^X}{X}\right)=-2
\left(\left(\frac{\partial_XW}{X^*}\right)^*+\frac{A_{\kappa_1}}{3}\right)\simeq
0,\eea upon ignoring the gravity mediated contribution of ${\cal
O}(m_{3/2})$.  Note that the equation of motion of $Y$ leads to the
cancellation between the two contributions to $B(M_\Phi)$, making
$B(M_\Phi)$ even smaller than ${\cal O}(A_{\kappa_1})$. As the $B$
parameter at $M_\Phi$ is negligible compared to $m_{\rm soft}$, its
low energy value is determined by  the RG running from $M_\Phi$ to
the weak scale. In our case, the messenger scale $M_\Phi$ is
required to be higher than the PQ scale $v_{PQ}\sim \sqrt{m_{\rm
soft}M_{GUT}}= {\cal O}(10^9-10^{10})$ GeV. As a result, a sizable
value of $B$ can be induced at the weak scale, giving
$\tan\beta=10\sim 20$, and furthermore its CP phase is automatically
aligned to the phase of gaugino masses.


With the PQ sector stabilized as above, we can identify the mass
eigenstates of the PQ sector fields and compute their mass
eigenvalues. First of all, the PQ sector provides a QCD axion having
a decay constant $v_{PQ}$ and thus a mass $m_a\sim f_\pi
m_\pi/v_{PQ}$, which corresponds mostly to the phase degree of
freedom of $X$. It contains also three real scalars with a mass
comparable to $m_{\rm soft}$, i.e. the saxion $x$ which is mostly
the modulus of $X$ and two others from $Y=y_1+iy_2$, and two
Majorana fermions $\tilde{a}_i$ ($i=1,2$) which form approximately a
Dirac axino $\tilde{a}=(\tilde{a}_1,\tilde{a}_2)$. It is then
straightforward to
 find
 \bea
 \label{mass}
m_{x} \simeq 2|m_X| ,\quad m_{y_1} \simeq m_{y_2}\simeq
m_{\tilde{a}}\simeq \sqrt{3}|m_X|, \eea where $m_X$ is given by
(\ref{mx}).

In summary, in our model the messenger scale of gauge mediation is
assumed to be higher than  the PQ scale, and then the
$U(1)_{PQ}$-breaking  fields $X$ and $Y$ are destabilized from the
origin due to the tachyonic  soft mass of $X$ and the scalar $A$
term  associated with the nonrenormalizable superpotential term
$\kappa_1X^3Y/6\Lambda$. These soft SUSY breaking terms of $X$ and
$Y$  are induced by the combined effects of gauge mediated SUSY
breaking and the Yukawa coupling $\lambda_XX\Psi\Psi^c$. The
supersymmetric scalar potential from the nonrenormalizable
superpotential prevents the runaway of $X$ and $Y$,  and stabilizes
them as
 \bea
\langle |X| \rangle &\sim&
\left(\frac{\lambda_X}{\kappa_1}\right)^{1/2}
\left(\ln\frac{M_\Phi}{\langle |X| \rangle}\right)^{1/4} \sqrt{m_{\rm
soft}\Lambda},\nonumber \\
\langle |Y| \rangle &\sim&
\frac{g_s^2\lambda_X}{32\pi^3}\left(\ln \frac{M_\Phi}{\langle |X| \rangle}\right)^{3/2}\langle
|X| \rangle, \eea where the messenger scale $M_\Phi> v_{PQ}\equiv
\langle |X|\rangle$ and $\Lambda$ is the cutoff scale of the model.
If we assume that $\lambda_X, \kappa_1$ and $\kappa_2$ are all of
order unity and $\Lambda \sim M_{\rm GUT} = 2 \times 10^{16}$ GeV,
while $M_\Phi$ is not so far above $v_{PQ}$,
 the mass scales of the model are
estimated as \bea
&& \mu \,\sim\,  m_{\rm soft} \,\sim\,  m_{PQ} \,=\, {\cal O}(10^2-10^3) \,\, {\rm GeV}, \nonumber \\
&& v_{PQ} \,\sim\,  \sqrt{m_{\rm soft} M_{\rm GUT}}\, =\, {\cal
O}(10^9-10^{10}) \,\, {\rm GeV},  \eea where $m_{PQ}$ stands for the
masses of the PQ sector fields (other than the QCD axion), including
the saxion and axino masses. The $B$ parameter at the messenger
scale is negligible compared to $m_{\rm soft}$, and therefore its
weak scale value is determined by the RG evolution below the
messenger scale, making its CP phase automatically aligned to the
phase of gaugino masses.


\section{Cosmology of the model}

The model described in Sec.~2 has a variety of interesting
cosmological implications. Because the PQ preserving field
configuration $X=Y=0$ is a local minimum of the effective potential
at high temperature $T\gg m_{\rm soft}$,
 it is a quite plausible
possibility that $X$ is settled down at the origin after the
primordial inflation. Then
 the early Universe experiences a late thermal
inflation \cite{Lyth:1995hj,CCK} before the PQ phase transition occurs,
 which might be
useful to eliminate (or dilute) potentially dangerous cosmological
relics such as light moduli or gravitinos.\footnote{Note that
a UV completion of the model within the framework of supergravity or
string theory might contain cosmologically harmful light moduli
causing the so-called moduli problem.}. Since this thermal inflation
will erase out any primordial baryon asymmetry, we need a
baryogenesis mechanism operating after thermal inflation is over. As
for the dark matter in our model, one can consider two possible
candidates, QCD axion with a mass $m_a\sim {f_\pi m_\pi}/{v_{PQ}}$
and light gravitino with a mass $m_{3/2}\sim {F}/{M_{Pl}}$.
 However the PQ scale of our model is determined as
$v_{PQ}\sim \sqrt{m_{\rm soft}M_{GUT}}\sim {\cal O}(10^9-10^{10})$
GeV, which might be too low to give a QCD axion constituting the
major fraction of the observed dark matter \cite{strongcp}. This
leads us to focus on the possibility of gravitino dark matter.

In the following, we briefly discuss the cosmological features of
our model, while leaving  more complete discussions for a separate
paper \cite{CCPS2}. As we will see, for the gravitino mass range
\bea m_{3/2}={\cal O}(100)\,\,{\rm keV},\eea
a right amount of gravitino dark matter can be produced after
thermal inflation, together with a successful Affleck-Dine
leptogenesis. As the gravitino mass is given by \bea m_{3/2}\sim
\frac{F}{M_{Pl}}\sim\frac{16\pi^2}{g^2}\frac{M}{M_{Pl}}m_{\rm
soft}\eea
 for
the SUSY breaking spurion $Z=M+F\theta^2$, this range of $m_{3/2}$
suggests that the messenger scale of gauge mediation, i.e.
$M_\Phi=\lambda_ZM$, which is presumed to be higher than the PQ
scale $v_{PQ}$, should be somewhat close to $v_{PQ}\sim
10^9-10^{10}$ GeV.

\subsection{Thermal inflation}

Near the origin, the finite-temperature effective potential of the
flat direction $|X|$ is given by \beq V(X) = V_0 + \left( \beta_X^2
T^2 - |m_X^2(0)| \right) |X|^2 + \cdots, \eeq where $V_0={\cal
O}(m_X^2 v_{PQ}^2)$ is the potential energy at the origin, which is
set to make the cosmological constant at true vacuum  vanish, and
$\beta_X$ comes from the couplings to thermal bath. Once the
Universe were in a radiation dominated period with $T> V_0^{1/4}$
after the primordial inflation is over, thermal inflation
begins at the temperature \beq T_\mathrm{b} \sim V_0^{1/4} \sim 10^6
\,\,{\rm GeV} \eeq
and ends when  $|X|$ is destabilized from the origin at the critical
temperature \beq T_\mathrm{c} = \frac{|m_X(0)|}{\beta_X}. \eeq


Soon after the end of thermal inflation, the Universe is dominated
for a while by the coherent oscillation of $|X|$ around its true
minimum $\langle |X|\rangle=v_{PQ}$, which eventually decays into
lighter particles to reheat the Universe. Since the saxion mass
$m_{x} =\mathcal{O}(10^{2}-10^3)$ GeV in our model (see (\ref{mass})), it can decay dominantly to the light Higgs boson
pair $h+h^*$ through the coupling of the form $\mu^2 hh^*\delta
x/v_{PQ}$, where $\delta x$ denotes the saxion fluctuation around
its vacuum.
 Assuming that $m_x$ is heavier than $2m_h$ for the light Higgs boson mass $m_h\simeq
120$ GeV, the decay rate of $|X|$ is estimated as \beq \Gamma_X \sim
\frac{1}{4 \pi} \frac{\mu^4}{m_{x} v_{PQ}^2},\eeq and then we find
the reheat temperature is given by \bea \label{TdX} T_\mathrm{RH}
&\equiv& \left( \frac{\pi^2}{15} g_*(T_{\mathrm{RH}}) \right)^{-1/4}
\Gamma_X^{1/2} M_\planck^{1/2}
\\
&\simeq& 1 \TeV \left( \frac{300 \GeV}{m_{x}} \right)^{1/2} \left(
\frac{\mu}{600 \GeV} \right)^2 \left( \frac{3 \times 10^9
\GeV}{v_{PQ}} \right), \eea where $g_*(T_\mathrm{RH}) \sim 100$ is
the number of light degrees of freedom at $T=T_{RH}$. The total
number of $e$-foldings of this thermal inflation is estimated to be
about $10$ and the dilution factor due to the entropy release in the
decay of $|X|$ is about $\mathcal{O}(10^{10})$. This would be large
enough to remove for instance the gravitinos produced before thermal
inflation \cite{Moroi:1993mb}.

Our model has two other oscillating scalar fields which are mostly
$y_1={\rm Re}(Y)$ and $y_2={\rm Im}(Y)$.
Although they have a mass comparable to $m_{x}$ (see (\ref{mass})), their energy densities are suppressed by $\langle
|Y|^2\rangle /\langle |X|^2\rangle ={\cal
O}(g_s^4\lambda_X^2/(8\pi^3)^2)$ compared to that of $|X|$, and
therefore they do not give a significant impact on the cosmological
evolution after thermal inflation.


\subsection{ Affleck-Dine leptogenesis}

Thermal inflation erases pre-existing baryon asymmetry. One may
think that an Affleck-Dine (AD) baryogenesis before thermal
inflation with a very large initial value of AD field can produce
enough baryon asymmetry which would survive after thermal inflation
\cite{de Gouvea:1997tn}. However it is known that the formation of
Q-balls makes it difficult to realize such scenario
\cite{Kasuya:2001tp}. Fortunately, our model can realize the
late-time AD leptogenesis proposed in Refs.
\cite{Stewart:1996ai,Jeong:2004hy,Kawasaki:2006py,Felder:2007iz,Kim:2008yu,Choi:2009qd,Park:2010qd}.

In order for the AD leptogenesis to work, the MSSM flat direction
$LH_u$ is required to have a nonzero value at certain stage. In our
case, this initial condition can be achieved as $LH_u$ has a
tachyonic soft mass-square $-m_{LH_u}^2$ in the limit $\mu=0$, so
unstable at the origin if the temperature drops below its critical
temperatures $T_{LH_u}=m_{LH_u}/\beta_{L H_u}\sim \sqrt{2} m_{LH_u}$ and $X$ is still staying at the
origin.
This requires
  \beq \label{con} T_\mathrm{c}=\frac{|m_X(0)|}{\beta_X} <
\sqrt{2} m_{LH_u}, \eeq and thus \bea \label{con1} \beta_X^2 \,> \,\left(
\frac{m_X(0)}{m_{LH_u}} \right)^2 \,\simeq\,
\frac{6\lambda_X^2}{\pi^2}\left(\frac{m_{\tilde{\Psi}}}
{m_{LH_u}}\right)^2\ln\left(\frac{M_\Phi}{v_{PQ}}\right),
\eea where we have used $m_X(0) \simeq 4 m_X(v_{PQ})$ together with
the result (\ref{mx}) for $m_X(v_{PQ})$. In our model, $\beta_X^2$
receive a contribution from  the exotic quark superfields
$\Psi,\Psi^c$, giving \bea \Delta
\beta_X^2=\frac{3}{4}\lambda_X^2.\eea It turns out that it is
difficult to satisfy (\ref{con1}) only with $\Delta \beta_X^2$ for
typical parameter values of our model. However this difficulty can
be easily avoided if the field $X$ couples to the right-handed
neutrinos $N$ to generate the seesaw scale \cite{kang}. Then there
will be an additional contribution to $\beta_X^2$ from the Yukawa
coupling $\lambda_NX NN$ \cite{Comelli:1996vm},
 \beq \beta_X^2 = \frac{1}{4} \left(
\sum_N \lambda_N^2 + 3\lambda_X^2 \right), \eeq
with which the condition (\ref{con1}) can be satisfied with a
reasonable value of $\lambda_N$.

Once the key condition  (\ref{con1}) for AD leptogenesis is
satisfied, the AD field $LH_u$ rolls down to nonzero value at the
temperature $T\sim m_{LH_u}$. If $T$ drops further down to $T_c$,
$|X|$ rolls away from the origin to generate nonzero $\mu$, and then
$LH_u$ gets a positive mass-square due to the contribution from
$\mu^2$. As a result, $LH_u$ rolls back to the origin with an
angular motion generated by CP-violating terms in the scalar
potential. The lepton asymmetry associated with the angular motion
of AD field is finally converted to baryon asymmetry through the
sphaleron process. The resulting baryon asymmetry at present is
estimated as \cite{Jeong:2004hy}
\bea \label{YBatpresent} \frac{n_\mathrm{B}}{s} \sim \frac{3}{8}
\frac{n_\mathrm{L}T_{\rm RH}}{n_xm_x}, \eea where $n_x\sim m_x
v_{PQ}^2$ is the saxion number density for coherently oscillating
saxion field $|X|$, and $n_L$ is the lepton number density
associated with the angular motion of the AD field. In fact, $n_L$
depends on many details of the full scalar potential, including the
terms associated with the lepton number violating neutrino mass term
in the superpotential. It depends for instance on the initial
displacement of the AD field from the origin,
curvature of the potential in angular direction, CP phase, e.t.c.
Using the results of
\cite{Jeong:2004hy,Felder:2007iz,Kim:2008yu,Park:2010qd}, we  find
that a value of $n_L={\cal O}(10^{11}-10^{12})\,\, {\rm GeV}^3$ can
be achieved under a reasonable assumption on the involved model
parameters,  and therefore the AD leptogenesis after thermal
inflation can produce the observed value of $n_B/s\sim 10^{-10}$
within the uncertainties in the involved parameters.

\subsection{Dark matter}

In our model the lightest supersymmetric particle (LSP) is the
gravitino, because all other supersymmetric particles including
axinos have mass of order $m_{\rm soft}$, and
 the gravitino mass is much smaller than $m_{\rm soft}$.
Thus light gravitino is the prime candidate of the dark matter.
On the other hand, thermal inflation also dilutes pre-existing
gravitino relics. After that, by decay of $|X|$, the Universe is
reheated with temperature $T_{\rm RH}\sim 1 {\rm TeV}$ as in Eq.~(\ref{TdX}).
At this temperature, most of MSSM fields are thermalized and will
produce light gravitinos. We can divide this process into two parts.
One is thermal (TH) production in which gravitinos are produced by
scatterings and decays of the MSSM fields in thermal bath. The other
is non-thermal (NTH) production in which gravitinos are produced
by out of equilibrium decays of the frozen relics such as the next
LSP which is the ordinary LSP (OLSP) in the MSSM sector or the axino.
The corresponding relic density of the gravitino can be represented by
\bea
\Omega_{3/2}h^2 = \Omega_{3/2}^{\rm TH}h^2 +
\Omega_{3/2}^{\rm NTH}h^2 \simeq 2.8\times 10^{4}
\left(\frac{m_{3/2}}{100\ \rm keV}\right) \left(Y_{3/2}^{\rm TH}
+ Y_{3/2}^{\rm NTH}\right) \eea 
where $Y^{\rm (N)TH}_{3/2} = n^{\rm (N)TH}
_{3/2}(T)/s(T)$ is the yield of the gravitino which is produced
by (non-)thermal process. At $T_{\rm RH}\sim 1 {\rm TeV}$, gravitinos
from thermal production can provide a right amount of cold dark
matter, $\Omega^{\rm TH}_{3/2}h^2 \simeq 0.1$, if the mass is \cite{Moroi:1993mb}
\bea m_{3/2} \sim 100\ {\rm  keV}. \eea
For the NTH production of the gravitinos, the contribution from the OLSP
decay is small enough in the above range of $m_{3/2}$, but we have to
pay attention to the production from the axino decays.
Although the axino couplings to the MSSM particles are suppressed by $1/v_{\rm PQ}$, they are still large enough to generate a significant axino abundance from the thermal bath. If the axino is stable, one needs $T_{\rm RH} \ll m_{\rm soft}$ in order to suppress its relic density sufficiently  \cite{chun00}. In our case of $T_{\rm RH}\sim m_{\rm soft}$, thermally generated axinos may produce a large number of gravitinos from their decays.  If axinos decay only to gravitinos, the non-thermal relic density of gravitinos turns out to be too large.
Thus, the axino decay to the gravitino must be suppressed.
For this,
let us now consider  the following axino mass range:
\bea  m_{\chi} + m_h < m_{\tilde a} < \mu  - m_h \eea where $\chi$ is the
OLSP. Then the dominant production and decay channels of the axino come from the
higgs($h$)-higgsino($\tilde{h}$)-axino($\tilde a$) coupling:
\bea
\int d^2\theta \frac{\kappa_2}{2\Lambda}X^2 H_u H_d = \frac{\mu}{v_{PQ}}h\tilde h\tilde a +
\cdots.\eea
That is, axinos are  produced thermally by the process
$\tilde h \to h \tilde a$, and then they decay mainly through   $\tilde a \to h\chi$.
Denoting the decay rate of higgsino to axino as $\Gamma(\tilde h \to h \tilde a) $, one finds the  thermal axino abundance as follows:
\bea \label{Yaxino}
 Y^{\rm TH}_{\axino} \approx {135 \zeta(3) \over 8 \pi^4 g_* }
 \left.{\Gamma(\tilde h \to h \tilde a) \over H}\right|_{T=\mu} \eea
where $g_*\sim 200$ is the relativistic degrees of freedom  and $H\approx 0.33 \sqrt{g_*} T^2/M_P$ is the Hubble parameter at the termperature $T$.  Then, the non-thermal abundance of the gravitino is
\bea \label{YNTH}
Y^{\rm NTH}_{3/2} = { \Gamma({\tilde a\to \psi_{3/2} \, a})  \over
\Gamma({\tilde a \to \chi h}) } Y^{\rm TH}_{\tilde a} \eea
in which partial decay rates of axinos are given by $\Gamma({\tilde a\to \psi_{3/2} \, a}) = m^5_{\tilde a}/ 96\pi m^2_{3/2} M_P^2$ \cite{chun94} with $\psi_{3/2}$ being the gravitino and $\Gamma({\tilde a \to \chi h}) = \theta^2 \mu^2 m_{\tilde a} /8\pi v_{\rm PQ}^2$. Here $\theta$ parameterizes the OLSP fraction in the Higgsino component. For our estimation, we will use  $\theta \sim m_Z s_W/\mu$
which is valid in the limit of the large Higgsino and small gaugino masses.
Combining (\ref{Yaxino}) and (\ref{YNTH}), we get
\bea
Y^{\rm NTH}_{3/2} \approx 1.5\times 10^{-7} \left(m_{\tilde a} \over 350\, \rm GeV\right)^4
\left(100\, {\rm keV} \over m_{3/2}\right)^2 \left( 600\,\rm GeV \over \mu\right)
 \eea
which shows that the non-thermal gravitino relic density can be safely neglected.
Finally let us remark that the axino decays well before the OLSP freezes out, that is,
$\Gamma({\tilde a \to \chi h}) > H(T_f)$ for the OLSP freeze-out temperature $T_f \lesssim 20\, \rm GeV$ for our choice of parameters.  Therefore, the OLSPs from the axino decay are thermalized.

\section{Conclusion}

In this paper, we have examined a model to generate the Higgs $\mu$
parameter with a spontaneously broken $U(1)_{PQ}$ symmetry in gauge
mediation scenario. The PQ sector of the model contains $U(1)_{PQ}$
breaking fields which have a Yukawa coupling to extra quarks. The
$U(1)_{PQ}$ breaking fields also have a nonrenormalizable
superpotential  suppressed by the cutoff scale $\Lambda$ which might
be identified as the Planck scale or the GUT scale. For the
messenger scale higher than the PQ scale,
the $U(1)_{PQ}$ breaking fields are destabilized at the origin due
to the soft SUSY breaking terms induced by the combined effects of
gauge mediated SUSY breaking and the Yukawa coupling to extra
quarks. They are then stabilized by the supersymmetric scalar
potential from nonrenormalizable superpotential  at an intermediate
scale $v_{PQ}\sim \sqrt{m_{\rm soft}\Lambda}$,  generating $\mu\sim
v_{PQ}^2/\Lambda \sim m_{\rm soft}$ in a natural manner. The $B$
parameter at the messenger scale is predicted to be negligible, and
therefore  $B$ at the weak scale is determined by the RG evolution
below the messenger scale.

The model has a variety of interesting cosmological features
associated with the PQ phase transition. In particular, a late
thermal inflation is a natural possibility, which would require a
late baryogenesis mechanism. We find that a successful Affleck-Dine
leptogenesis after thermal inflation can be implemented  within the
model. We also find that a right amount of gravitino dark matter can
be produced after thermal inflation when $v_{PQ}={\cal O}(10^9 -
10^{10})$ GeV and $m_{3/2}={\cal O}(100)$ keV, for which the
messenger scale of gauge mediation is required to be not far above
$v_{PQ}$. More complete discussion of the cosmological aspects of
the model will be presented elsewhere.

\section*{Acknowledgments}
KC and CSS are supported by the KRF Grants funded by the Korean Government (KRF-2008-314-C00064 and KRF-2007-341-C00010) and the KOSEF Grant funded by the Korean Government (No. 2009-0080844). 
EJC is supported by Korea Neutrino Research Center through National Research Foundation of Korea Grant (2009-0083526). 
HK is supported by KRF-2008-313-C00162.

\end{document}